\documentclass[a4paper]{article}

\usepackage{fullpage} 
\usepackage{pstricks}
\usepackage{multicol}
\usepackage{graphicx,graphics}
\usepackage{movie15}
\usepackage{hyperref}
\usepackage[version=3]{mhchem}
\usepackage{amsmath} 
\usepackage{amsfonts}
\usepackage{color}
\usepackage{soul}
\usepackage[margin=1.7cm,top=3cm,bottom=2.8cm,centering]{geometry}

\definecolor{rosepale}{rgb}{1.0, 0.7, 1.0}

\newcommand{\be}{\begin{equation}}
\newcommand{\ee}{\end{equation}}
\newcommand{\bea}{\begin{eqnarray}}
\newcommand{\eea}{\end{eqnarray}}
\newcommand\egal{&\!\!\!\!=\!\!\!\!&}

\def\la{\langle}
\def\ra{\rangle}

\title{Twelve competing ways to recover the Michaelis-Menten equation reveal the alternative approaches to steady state modeling \footnote{Reference: This manuscript is an expanded version of the article: Michel, D. \& Ruelle, P. 2013. Seven competing ways to recover the Michaelis-Menten equation reveal the alternative approaches to steady state modeling J. Math. Chem. 51, 2271-2284.
DOI 10.1007/s10910-013-0237-5}}

\author{Denis Michel$ ^{a} $ and Philippe Ruelle$ ^{b} $ \\
\\
\begin{small} $ ^{a} $ Universite de Rennes1-IRSET. Campus de Beaulieu. 35000 Rennes France. denis.michel@live.fr \end{small} \\
\begin{small} $ ^{b} $Universit\'e catholique de Louvain - UCL. Institut de Recherche en Math\'ematique et Physique.\end{small} \\ \begin{small} IRMP. Chemin du Cyclotron, 2. B-1348 Louvain-la-Neuve, Belgium. philippe.ruelle@uclouvain.be \end{small}\\}

\date{} 
\begin{document}
\maketitle

\begin{multicols}{2}

\textbf{The Michaelis-Menten enzymatic reaction is sufficient to perceive many subtleties of network modeling, including the concentration and time scales separations, the formal equivalence between bulk phase and single-molecule approaches, or the relationships between single-cycle transient probabilities and steady state rates. Twelve methods proposed by different authors and yielding the same famous Michaelis-Menten equation, are selected here to illustrate the kinetic and probabilistic use of rate constants and to review basic techniques for handling them. Finally, the general rate of an ordered multistep reaction, of which the Michaelis-Menten reaction is a particular case, is deduced from a Markovian approach.} \\

\section{Preliminary tools: the helpful scale separations}

The approximation of concentration and time scales separations is often reasonable in cellular biochemistry, but some discernment is necessary for its proper application, in particular to define pseudo-first order constants and decide which reactions can be considered as in quasi-equilibrium compared to others.

\subsection{Concentration scale separation}
The wide differences of molecular concentrations in the cell greatly facilitate network modeling. The concentration of the more concentrated reactant, generally called the ligand or the substrate in enzymology, can be associated to second-order constants to give so-called pseudo-first order constants. This approximation strongly simplifies elementary treatments, for example to define hyperbolic saturation functions  through equating the concentrations of total and free ligand. To apply the concentration scale separation, it is important to decide which reactant should be fused to the second-order constant. Depending on the cases, the same molecule can behave as either the leading macromolecule or as the ligand. This is the case, for example, for a transcription factor, say the estrogen receptor (ER), activated by the estrogen hormone (E2) and then capable of binding to a given unique gene (G) from the X chromosome. Even if ERs are not very numerous in the cell, they are however much more abundant than the single gene. Hence, ER can be considered as a diffusible ligand whose concentration varies slowly compared to the dynamics of its interaction with the gene. Conversely, when studying the activation of ER by E2, the ligand is now E2, which should be integrated in pseudo-first order rates of ER state changes. If the binding of E2 to ER and the binding of ER to G are to be mixed in the same model, a new approximation intervenes: the time scale separation. 

\subsection{Time scale separation}

In the example introduced above, the interaction between E2 and ER can be considered as more dynamic than that occuring between ER and G, that is itself dynamic compared to the time scale of gene expression, transcription and translation. Time scale separation is particularly important to obtain smooth graded interactions between a ligand and a very unique binding site \cite{michel}. This approximation allows kinetic and equilibrium constants to coexist in the same equation, as long proposed \cite{Cha}. Once built, the first-order network can be treated through different methods yiedling equivalent results. A survey of some of them is proposed below using a founder example of historical importance: the Michaelis-Menten (MM) enzymatic reaction. In these methods, rate constants are envisioned as frequencies corresponding to the inverse of mean waiting times. Indeed, molecular events are primarily dictated by waiting times whereas the transitions are themselves considered as instantaneous. Before examining the MM reaction, it is first necessary to introduce elementary recipes mixing kinetic and probabilistic thinking \cite{Ninio1986,Ninio1987} and which can be addressed using the simple questions of Fig.1.

\section{Shortcuts to steady state modeling}

Fundamental kinetic rules are illustrated by the three simple reactions shown in Fig.1. For (a) and (b), what is the rate constant of the global transition from A to B and what is the mean lifetime in state A ? For (c), what is the probability for A to shift to B rather than to C and what is the mean lifetime in state A ?

\begin{center}
\includegraphics[width=6cm]{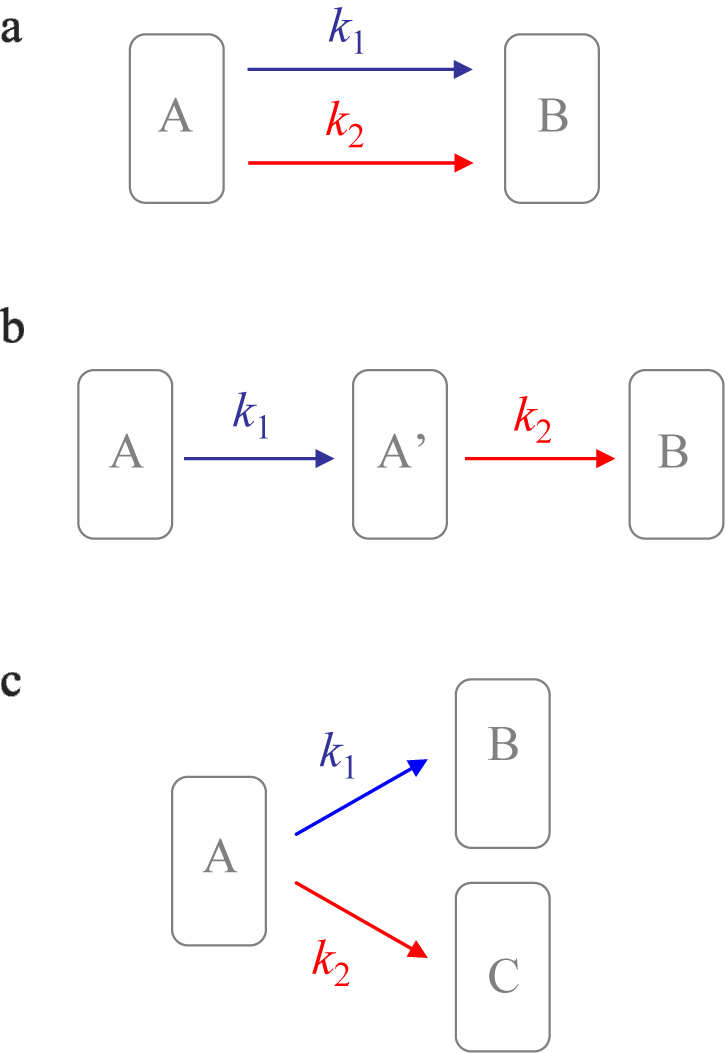} \\
\end{center}
\begin{small} \textbf{Figure 1.} Basic rules of direct rate constant manipulation.  \end{small}\\

In Fig.1a, there are two ways to leave A so that the global rate is the sum of the individual rates $ k= k_{1} + k_{2}  $. This intuitive result is related to a property of the exponential distribution $ \xi (\theta ) $. If $X=\xi(\lambda)$ and $Y=\xi(\mu)$ are two independent random variables with exponential law, then $\min(X,Y)=\xi(\lambda+\mu)$. Indeed, $\xi(\theta)$ is characterized by its tail $P(\xi\geq x)=\textup{e}^{-\theta x}$. Hence $P(\min(X,Y)\geq x)=P(X\geq x $ and $ Y\geq x) $ = $ P(X\geq x)P(Y\geq x) = \textup{e}^{-\lambda x}\textup{e}^{-\mu x} = \textup{e}^{-(\lambda + \mu) x} $. 

The rate of the global transition of Fig.1b is obviously lower than that of Fig.1a since there are two waiting times from A and B. The mean time to reach B from A is $ 1/k_{1} + 1/k_{2} $ and the mean lifetime of A is $ 1/k_{1} $. Hence, in case of continuous supply with A, the frequency of the arrivals to B is the reciprocal of its mean time: $ k= k_{1}k_{2}/(k_{1} + k_{2}) $. 

For Fig.1c, the probability to obtain B is $ k_{1}/(k_{1} + k_{2}) $ and the mean lifetime in state A is the same as for Fig.1a: $ 1/(k_{1} + k_{2}) $. These results are the very basic ingredients of network modeling and will be used in the following section.

\section{Twelve ways to the Michaelis-Menten reaction rate}

The 2-step Michaelis-Menten enzyme reaction scheme represented in Fig.2 has become an unavoidable chapter of enzymology textbooks and a masterpiece of biochemistry courses.

\begin{center}
\includegraphics[width=6cm]{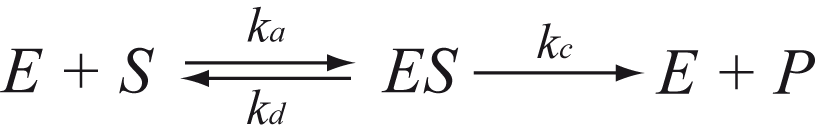} \\
\end{center}
\begin{small} \textbf{Figure 2.} The famous Michaelis-Menten enzymatic reaction, where an enzyme $ E $ catalyses the conversion of a substrate $ S $ into a product $ P $. The final reaction of rate $ k_{c} $ is considered as irreversible. \end{small}\\
\\
Interestingly, a sort of game has been established by different authors to test their alternative methods of treatment using the MM reaction. Some of them are of great pedagogical interest and will be listed below, but let us first consider the pioneer work of Michaelis and Menten.

\subsection{Michaelis-Menten: the quasi-equilibrium assumption}
The names of Michaelis and Menten \cite{Michaelis} are associated with both the hyperbolic equilibrium fraction of saturation $ Y=[S]/(K_{d}+[S]) $ and the elementary steady state enzymatic reaction rate $ k=k_{c}[S]/(K_{M}+[S]) $ where $ K_{M} $ is precisely known as the Michaelis constant. But in fact, if the founding contribution of Michaelis and Menten to biochemical modeling cannot be denied, they did not discover these two fundamental bases of systems biology. The Michaelis-Menten equilibrium hyperbola, also known as the isotherm of Langmuir, was described by Jean-Baptiste Biot \cite{Biot} and the Michaelis constant was introduced by Briggs and Haldane \cite{Briggs}. The name of Michaelis was perhaps given to the Michaelis constant in honour of his previous scientific contribution. Nevertheless, the approach of Michaelis-Menten remains interesting because they seemed to assume, non-explicitely or unintentionally, the principle of time scale separation. As explained by Briggs and Haldane, Michaelis and Menten considered that the kinetics of noncovalent interaction between the enzyme and the substrate is much more rapid than that of the catalytic reaction, so that they used the dissociation equilibrium constant instead of the Michaelis constant introduced later. Further confusing the situation, Michaelis and Menten did not write their constant $ K_{d} $ but $ K_{s} $ \cite{Michaelis}, which should be interpreted as the $ K_{d} $ for the substrate ($ S $) \cite{Briggs}. 

The method of Michaelis-Menten can be described as follows. Given that catalysis concerns only the fraction of enzyme bound to the substrate, the rate of product synthesis ($ k $) normalized by the total amount of enzyme $ [E]_{tot} $ is 

\begin{subequations} \label{E:gp}
\begin{equation} 
k=k_{c}\frac{[ES]}{[E]_{tot}}=k_{c}\frac{[ES]}{[E]+[ES]}. 
\end{equation} 

where $ [E] $ is the concentration of the free enzyme. Considering that the catalytic reaction is infrequent compared to the binding and dissociation reactions, the substrate and the enzyme can be assumed to be in equilibrium and the equilibrium  constant can be written as a ratio of concentrations,
\begin{equation}  
K_{d}= \frac{[E][S]}{[ES]}.
\end{equation}

Combining these two equations yields
\begin{equation} 
k=k_{c}\frac{[S]}{K_{d}+[S]}. 
\end{equation} 

\end{subequations} 

\subsection{Briggs and Haldane: the steady-state assumption}
In the classical scheme represented in Fig.2, the catalytic transition is represented by a one-way arrow. The system is thus said to be micro-irreversible, which means that it is out of equilibrium and can be sustained only through to a continuous refuelling with fresh substrate. An in depth analysis of the legitimacy of the quasi-steady state approximation is proposed in \cite{Dhatt}. The fraction of enzyme bound to the substrate is the steady state resultant of its formation and its disappearance according to \cite{Briggs}
\begin{subequations}
\begin{equation} 
k_{a}[E][S]=(k_{d}+k_{c})[ES]. 
\end{equation} 
Therefore we have 
\begin{equation} 
\frac{[E]}{[ES]}=\frac{k_{d}+k_{c}}{k_{a}[S]}, 
\end{equation} 
so that
\begin{equation} 
k=k_{c}\frac{[ES]}{[E]+[ES]}=\frac{k_{c}k_{a}[S]}{k_{d}+k_{c}+k_{a}[S]}=k_{c}\frac{[S]}{K_{M}+[S]},
\end{equation} 
where
\begin{equation} 
K_{M}=(k_{d}+k_{c})/k_{a}.
\end{equation} 
\end{subequations} 

\subsection{The first-order network of King and Altman}

The method of King and Altman \cite{King-Altman} was a precursor of modern single-molecule approaches in that it is basically probabilistic. Let us recall its rules. 
\begin{itemize}
\item[(i)] All rates should be first- or pseudo-first order ($ s^{-1} $). In the scheme of Fig.2, there is a second-order constant: $ k_{a} $ ($ M^{-1} s^{-1} $). As explained in Section 1, by virtue of the excess of substrate relative to the enzyme, a pseudo-first order constant can be defined $ k^{*}_{a}= k_{a}[S] $. 
\item[(ii)] A single kind of molecule should be considered in the different nodes of the network: the leading molecule is in the present case the enzyme. 
\item[(iii)] Finally, every state transition must be associated to a single arrow, possibly gathering several channels. 
\end{itemize}

The MM scheme of Fig.3 is a simplistic network made of only two nodes. Since there are two ways to convert $ [ES] $ into $ [E] $, it is necessary to give to this transition the rate obtained by summing the individual rates (as for Fig.1a). As a result, one obtains the first order network schematized in Fig.3. 

\begin{center}
\includegraphics[width=2.7cm]{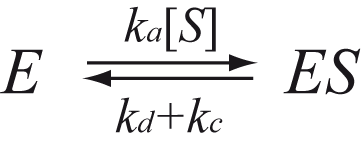} \\
\end{center}
\begin{small} \textbf{Figure 3.}  The MM reaction redrawn as a 2-node network in the manner of King and Altman. \end{small}\\

This scheme shows that a single enzyme can exist only in two forms. The probability of each form is straightforwardly defined by a King and Altman-like graphical method:
\begin{subequations} \label{E:gp}
\begin{equation} 
P_{ES}= \frac{\rightarrow }{\leftarrow +\rightarrow } = \frac{k_{a}[S]}{k_{d}+k_{c}+k_{a}[S]}, 
\end{equation} 
and
\begin{equation} 
P_{E}=\frac{\leftarrow }{\leftarrow +\rightarrow } = \frac{k_{d}+k_{c}}{k_{d}+k_{c}+k_{a}[S]},
\end{equation} 
so that the rate is simply
\begin{equation} 
k=k_{c}P_{ES}=\frac{k_{c}k_{a}[S]}{k_{c}+k_{d}+k_{a}[S]}. 
\end{equation} 
\end{subequations} 

\subsection{The frequency of successful enzyme cycles}

The total time $\la T \ra$ necessary for a single enzyme to achieve the reaction of Fig.2 is the sum of two residence times: that of $ E $ waiting for catching a substrate molecule and then of $ ES $, waiting for either a reaction or a dissociation \cite{Ninio1986}. The first one is simply $ 1/k_{a}[S] $ and the second one is $ 1/(k_{d}+k_{c}) $ (see the question of Fig.1c), so that
\begin{equation}
\la T \ra = \frac{1}{k_{a}[S]} + \frac{1}{k_{d}+k_{c}}. 
\end{equation}

The turnover frequency is the reciprocal of this time, $ \nu = 1/\la T \ra $, but all the turnovers are not successful since $ [ES] $ can simply dissociate into $ E + S $ instead of giving $ E + P $. The probability of this latter possibility is $ P(ES \rightarrow P)= k_{c}/(k_{d}+k_{c}) $ (see Fig.1c), giving the final expression
\begin{equation} 
k = \dfrac{1}{\la T \ra} \left(\frac{k_{c}}{k_{d}+k_{c}}\right) =\frac{k_{c}k_{a}[S]}{k_{c}+k_{d}+k_{a}[S]}. 
\end{equation}

\subsection{The frequency of non-abortive cycles}

This is the pessimistic counterpart of the previous approach, but which works as well \cite{Qian}. The probability that $ES$ merely dissociates instead of reacting is the complementary to the previous one $ P(ES \rightarrow E+S)= k_{d}/(k_{d}+k_{c}) $ and in the case, the whole chain is reinitiated. This situation can be explicitly transcribed into
\begin{subequations}
\begin{equation} 
\la T \ra = \frac{1}{k_{a}[S]} + \frac{1}{k_{d}+k_{c}}+ \frac{k_{d}}{k_{d}+k_{c}} \la T \ra.
\end{equation}
It implies, after some rearrangements,
\begin{equation} 
k = \frac{1}{\la T \ra} = \frac{k_{c} k_{a}[S]}{k_{c}+k_{d}+k_{a}[S]}. 
\end{equation} 
\end{subequations} 

\subsection{Sum of direct conversion times}

This powerful method is inspired from \cite{Fisher,Knorre,Malygin}. In the steady state, the rate of product formation $ k $ is the inverse of the sum of the forward conversion times for all enzyme states, which can be written as
\begin{equation} 
k = \frac{1}{\sum_i T^{\rightarrow}_{i}},
\end{equation}
where $T^{\rightarrow}_{i} $ is the time necessary to directly reach the objective (release of the product) when starting from the enzyme state $ E_{i} $. For the Michaelis-Menten reaction, only two states of the enzyme are to be considered: $ E $ waiting for a substrate and $ ES $ waiting for reacting. For the latter, the calculation is immediate,
\begin{equation} 
T^{\rightarrow}_{ES} = \frac{1}{k_{c}}.
\end{equation}

For $E$, the forward rate is $k_{a}[S]$ conditioned by the probability that when bound to $S$, the reaction proceeds rather than the substrate dissociates, which is, as defined previously, equal to $k_{c}/(k_{d}+k_{c})$. Hence, one has
\begin{equation} 
T^{\rightarrow}_{ES} + T^{\rightarrow}_{E} = \frac{1}{k_{c}} + \frac{1}{k_{a}[S]\frac{k_{c}}{k_{d}+k_{c}}},
\end{equation}
which implies
\begin{equation} 
k = \frac{1}{T^{\rightarrow}_{ES} + T^{\rightarrow}_{E}} = \frac{k_{c}k_{a}[S]}{k_{c}+k_{d}+k_{a}[S]}. \end{equation}

\subsection{The point of view of a single substrate molecule}
For a given substrate molecule, the binding of a diffusing enzyme appears as in the scheme

\begin{center}
\includegraphics[width=5 cm]{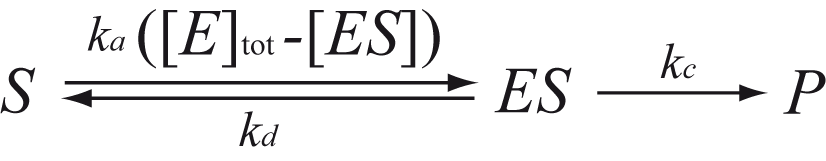} \\
\end{center}
The enzyme concentration included in the pseudo-first order binding constant is of course that of the free enzyme, which is itself determined by the population of substrate molecules which partly sequestrates it: $ [E]=[E]_{tot}-[ES] $. The steady state value of $ [ES] $ is given by
$$ k_{a}[S]([E]_{tot}-[ES])= (k_{d}+k_{c})[ES] $$
which gives
$$ [ES]= \dfrac{k_{a}[E]_{tot}[S]}{k_{d}+k_{c}+k_{a}[S]} $$
Introducing this value in the reaction rate $ v=k_{c}[ES] $ gives the final equation.

\subsection{The free enzyme activity}

Using the same probabilistic treatment as previously, the global reaction rate is

$$ v=  k_{a}[S][E] \dfrac{k_{c}}{k_{c}+k_{d}} = \dfrac{k_{c}}{K_{M}}[S][E] $$ where the concentration of the free enzyme, not sequestrated by other substrates, is 
$$ [E]=[E]_{tot} \dfrac{K_{M}}{K_{M}+[S]} $$ which gives, when reintroduced in the preceding rate, the correct final formula.

\subsection{The mean lifetime approach}

As illustrated in the bottom panel of Fig.1, rate constants can also be used to directly give mean lifetimes. In the elementary MM enzymatic scheme, this simple recipe gives
\begin{subequations}
\begin{equation} T(E)=1/(k_{a}[S])\end{equation}
and 
\begin{equation} T(ES)=1/(k_{d}+k_{c})\end{equation}
so that
\begin{equation} 
v=k_{c}\dfrac{T(ES)}{T(E)+T(ES)} =\frac{k_{c}k_{a}[S]}{k_{c}+k_{d}+k_{a}[S]}
\end{equation}
\end{subequations}

\subsection{Transient approach}

Interestingly, all the methods described above avoid transient treatments, but the mean cycling time obtained in this way expectedly gives the same results. At the single-enzyme level, a single-MM reaction follows a biexponential cycle \cite{Xie1998}. The MM behaviour can be recovered as the reciprocal of the mean time of a single cycle between the enzyme state waiting for a substrate ($ E_{0} $) and the enzyme state that released the first product ($ E_{1} $). The transient behaviour of this cycle is obtained by solving the following system
\begin{subequations}
\bea
\frac{\textup{d}P_{E_{0}}(t)}{{\rm d}t} \egal k_{d}P_{ES}(t) -k_{a}[S]P_{E_{0}}(t),\\
\frac{\textup{d}P_{ES}(t)}{{\rm d}t} \egal k_{a}[S]P_{E_{0}}(t)-(k_{d}+k_{c})P_{ES}(t),\\
\frac{\textup{d}P_{E_{1}}(t)}{{\rm d}t} \egal k_{c}P_{ES}(t).
\eea
At every time, a single enzyme can take only one of these three states so that
\begin{equation} 
P_{E_{0}}+P_{ES}+P_{E_{1}} = 1. 
\end{equation}
\end{subequations}

With the initial condition $P_{E_{0}}(0)=1$, this system predicts that the probability of cycle achievement at time $t$ approaches 1 according to
\begin{subequations} 
\begin{equation}
P_{E_1}(t) = 1-\textup{e}^{-\mu t}\left (\frac{\mu }{\lambda }\sinh \lambda t+\cosh \lambda t \right), 
\label{PE1}
\end{equation}
where
\begin{equation} 
\mu = (k_{d}+k_{c}+k_{a}[S])/2\,, \qquad \lambda=\sqrt{\mu ^{2}-k_{c}k_{a}[S]}.
\end{equation}
\end{subequations} 

The first moment of the distribution (\ref{PE1}) gives the mean time of a single cycle,
\begin{equation}
\la T \ra = \int_{0}^{\infty} t\, P_{E_1}(t)\, {\rm d}t = \frac{2\mu}{\mu^{2}-\lambda^{2}} = \frac{k_{d}+k_{c}+k_{a}[S]}{k_{c}k_{a}[S]} 
\end{equation}
whose inverse is the average product formation rate.

\subsection{Markovian modeling}

A brute force approach is to define the general equation valid for any $ n $-step enzymatic reaction and then to simplify it for the case of the 2-step MM reaction. Indeed, the reaction of Michaelis-Menten is a minimal version of the heterogeneous linear walk represented in Fig.4 and involved in many research areas beside this particular context \cite{Doussal,Zhou,Codling}. Note that to obtain such a chain, it is sometimes necessary to prune some branches in case of disordered events (see an example in Appendix 5.1). This $ n $-step chain is micro-irreversible because $ k^{-}_{n} =0 $. This general chain can be used as a common frame adaptable to many different enzymatic reaction schemes, simply by replacing the first-order constants by appropriate pseudo-first order ones.

\begin{center}
\includegraphics[width=8 cm]{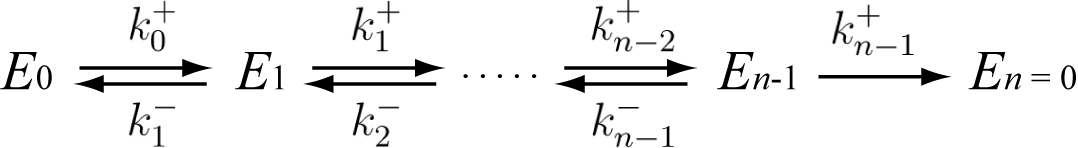} \\
\end{center}
\begin{small} \textbf{Figure 4.} First order chain where the different nodes represent different states of the same enzyme (conformation, post-translational modification, substrate(s) ligation(s) complexes etc.). The kinetic constants associated to each transition are labelled with + or - depending on whether they are forward or backward transitions and the indices refer to the starting state of the enzyme. It is important to note that the enzyme is not consumed during the reaction but is simply recycled ($ E_{n}=E_{0} $). \end{small}\\

 The mean arrival time of this generic reaction is (see Appendix 5.2 for a derivation of this expression)
\begin{equation} 
\la T \ra =\sum_{\ell=0}^{n-1}\sum_{i=0}^{n-\ell-1}\frac{1}{k^{+}_{i}}\prod_{j=i+1}^{\ell+i}\frac{k^{-}_{j}}{k^{+}_{j}}.
\label{T}
\end{equation}

This impressive formula dramatically simplifies for the Michaelis-Menten reaction, reduced to three constants $ k_{0}^{+} = k_{a}[S] $, $ k_{1}^{+} = k_{c}$ and $ k_{1}^{-} = k_{d} $,
\begin{equation} 
\la T \ra = \frac{1}{k_{a}[S]} + \frac{1}{k_{c}} + \frac{k_{d}}{k_{a}[S] k_{c}}.
\end{equation}

Considering that a single enzyme is recycled with a period $\la T \ra$ the frequency of the successive state conversion cycles corresponds to the reaction rate
\begin{equation} 
k = \frac{1}{\la T \ra} = \frac{k_{c}k_{a}[S]}{k_{c}+k_{d}+k_{a}[S]}.
\end{equation}

\subsection{The exact Michaelis-Menten equation}

The Michaelis-Menten equation can be defined in the general case, without any approximation about the relative concentrations of $ S $ and $ E $. It should be noted that all the methods listed above use the approximation that the substrate is much more abundant than the enzyme, so that the free and total concentrations of the substrate are considered equal. Although generally acceptable, this approximation implies that the substrate-dependent enzymatic velocity curves are false near the origin, at low substrate concentration. Hence, let us reformulate the exact Michaelis reaction formula holding for all substrate concentrations. The steady-state invariance of $ ES $ can be written

\begin{subequations} \label{E:gp}
\begin{equation} \dfrac{d+c}{a}=K= \dfrac{[E][S]}{[ES]} = \dfrac{([E]_{T}-[ES])([S]_{T}-[ES])}{[ES]} \end{equation}\label{E:gp1}
that is a quadratic equation

\begin{equation} [ES]^{2}- (K+[E]_{T}+[S]_{T})[ES]+[E]_{T}[S]_{T}=0 \end{equation}\label{E:gp2}

from which $ [ES] $ can be defined using the total concentrations of its components. Using this result and supposing that the enzyme has a single substrate and is not sequestrated by other molecules in the cell, the general Michaelis-Menten reaction is
\begin{equation} \begin{split} v=&c[ES]=\frac{c}{2}(K+[E]_{T}+[S]_{T} \\&-\sqrt{(K+[E]_{T}+[S]_{T})^{2}-4[E]_{T}[S]_{T}} )  \end{split} \end{equation}\label{E:gp2}
\end{subequations}

This exact equation can be simplified for a moderate sequestration. If $ [ES] << K+[E]_{T}+[S]_{T} $, the squared term in Eq.(18c) can be neglected and the solution is simply

\begin{subequations} \label{E:gp}
\begin{equation} [ES]= \dfrac{[E]_{T}[S]_{T}}{K+[E]_{T}+[S]_{T}} \end{equation}\label{E:gp1}
which itself simplifies, when $ [E]_{T}<<K,[S]_{T} $, into
\begin{equation} [ES]= \dfrac{[E]_{T}[S]_{T}}{K+[S]_{T}} \end{equation}\label{E:gp2}
that is the classical equation. In the atypical case of very low substrate $ [S]_{T}<<K,[E]_{T} $,
\begin{equation} [ES]= \dfrac{[E]_{T}[S]_{T}}{K+[E]_{T}} \end{equation}\label{E:gp3}
and finally in absence of any sequestration, when the interactions are very transient ($[S]_{T},[E]_{T} << K $)
\begin{equation} [ES]= \dfrac{1}{K}[E]_{T}[S]_{T} \end{equation}\label{E:gp3}
\end{subequations}
Fig.5 shows the comparison between the general Eq.(18c) and traditional equation.
\begin{center}
\includegraphics[width=8cm]{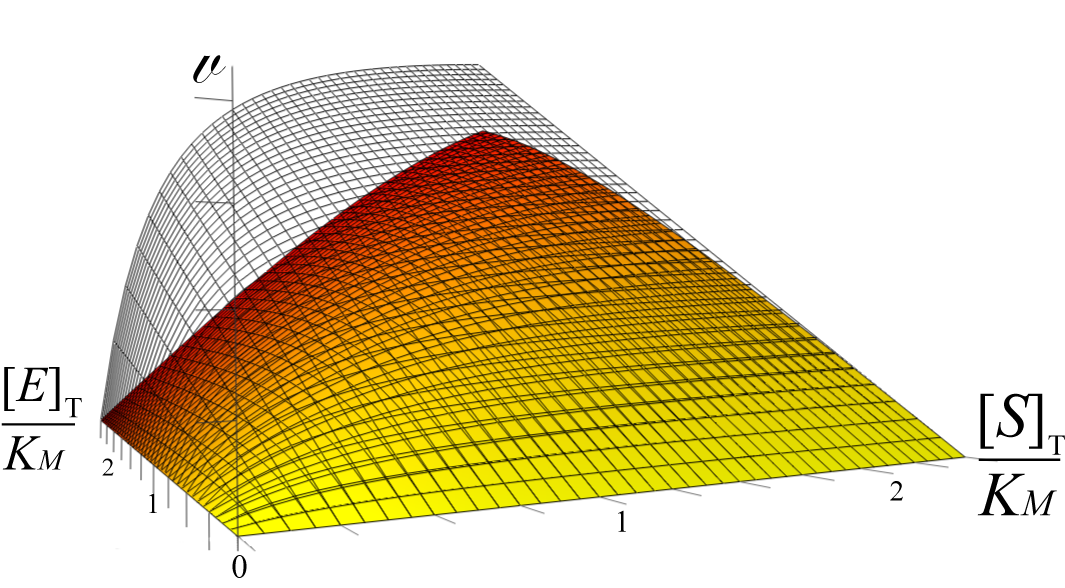} \\
\end{center}
\begin{small} \textbf{Figure 5.} Comparison of Michaelis Menten curves with $ K=0.5 $ between the classical approximation of saturating substrate (uncolored upper surface) and without approximation (colored lower surface, drawn to Eq.(18c)). The classical Michaelis-Menten treatement overestimates velocities at low substrate concentration and the linear dependence on the enzyme at fixed substrate concentration is no longer valid. The two surfaces converge at high substrate concentration. \end{small}\\
\newline

\section{Conclusion}

The previous section includes twelve methods, of which only the first one does not correspond to a steady state situation. Somewhat ironically the so-called Michaelis-Menten reaction has not been properly described by Michaelis-Menten. Alternative methods exist to recover the steady state MM equation in addition to those reviewed here, including for example the chemical master equation \cite{Qian2}, but the methods listed above provide the main recipes for modeling first order biochemical networks, which are the basics of systems biology.

\section{Appendices}

\subsection{Branch pruning to generate a linear chain}

The reduction of disordered cycles to ordered transitions has been addressed for example  in \cite{Ninio1987,Malygin}. The most general treatment of a linear chain with one input and one output and including a branched module, leads to a Michaelian kinetics with respect to the distribution of the reaction time between the input and output, irrespective of the complexity of the intervening module.

\begin{center}
\includegraphics[width=7cm]{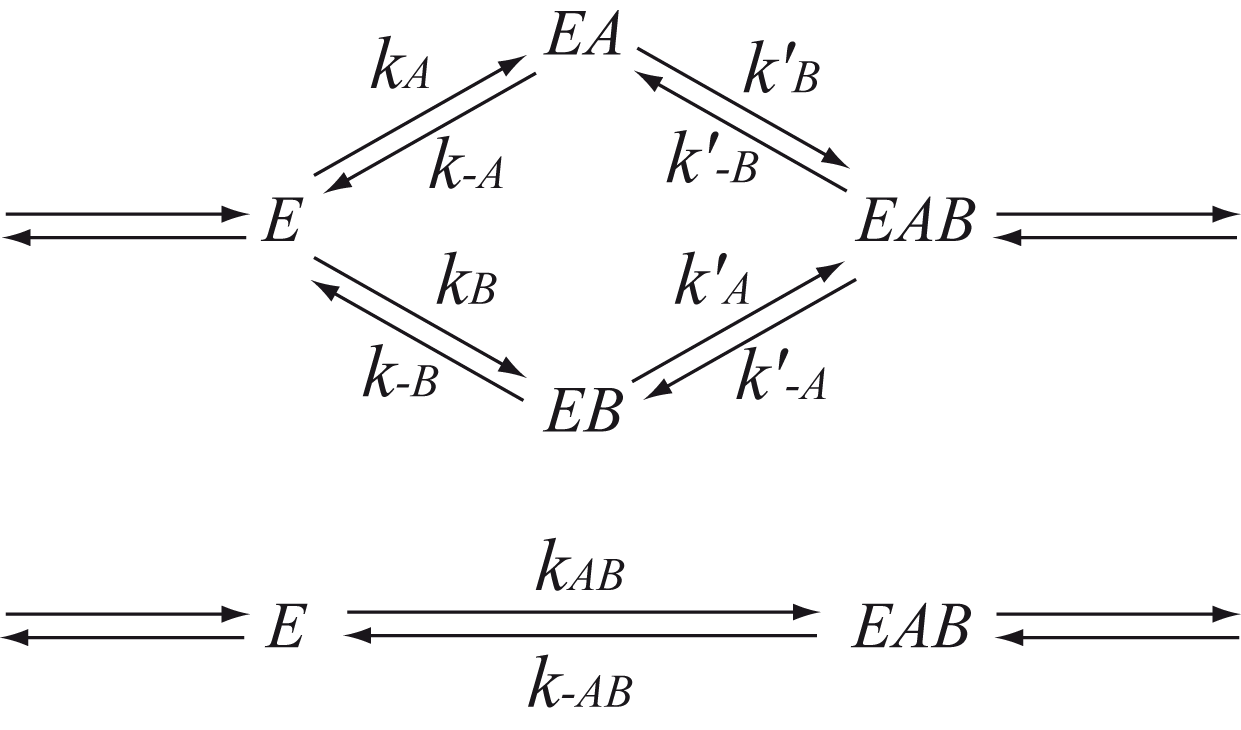} \\
\end{center}
\begin{small} \textbf{Figure 6.} How to compress lateral alternative states (top scheme) into an unbranched ordered chain (bottom scheme). \end{small}\\

\noindent
In a simplified case, if the fraction of time spent by the enzyme in the forms $ EA $ and $ EB $ (Fig.6), is negligible compared to its total mean cycling time, these alternative forms can be eliminated and the disordered individual rates can be replaced by ordered rates according to
\begin{subequations}
\bea  k_{AB} \egal k_{A}\, \frac{k'_{B}}{k'_{B}+k^{}_{-A}} + k_{B}\, \frac{k'_{A}}{k'_{A}+k^{}_{-B}}, \\
k_{-AB} \egal k'_{-B}\, \frac{k_{-A}}{k^{}_{-A} + k'_{B}} + k'_{-A}\, \frac{k_{-B}}{k^{}_{-B}+k'_{A}}.
\eea
\end{subequations} 

In addition, postulating that the walk is random implies that the transitions $ k_{A} $ and $ k_{B} $ do not influence each other. For example, if they are the pseudo-first order rates of ligation to two substrates $ A $ and $ B $, the binding of one substrate has no influence on the binding of the other one. In this case, $ k^{}_{A} = k'_{A} $ and $ k^{}_{B} = k'_{B} $ and the global rates are
\begin{subequations}
\begin{equation}  
k_{AB} = \frac{k_{A}k_{B}}{R}\,, \qquad k_{-AB} = \frac{k_{-A}k_{-B}}{R},
\end{equation}
with
\begin{equation} 
R = \frac{(k_{A}+k_{-B})(k_{-A}+k_{B})}{k_{A}+k_{B}+k_{-A}+k_{-B}}.
\end{equation}
\end{subequations} 

\subsection{Mean completion time for a general chain}
All of the methods listed above appear roughly equivalent when dealing with the elementary reaction of Fig.1, but are not equally extensible to more complex schemes. For example, the mean lifetime method quickly becomes impractical when the number of states of the enzyme increases. By contrast, certain methods can be easily implemented, like the King and Altmann method \cite{Beard} and the Markov modeling, as developped below. Eq. (\ref{T}) can be rigorously obtained through a continuous time Markovian modeling. In the following formulation, the probabilities of the different enzymatic states are written $P_i(t)$, $i=0,1,\ldots,n$, and correspond to the amount of final state $i$ at time $t$. Their evolution is determined by a linear differential system,
\be
{{\rm d} \over {\rm d}t}\,P_i(t) = \sum_{j=0}^{n} \: \hat A_{ij} \, P_j(t)\,, \qquad i=0,1,\ldots,n,
\ee
with the initial condition $P_i(0) = \delta_{i,0}$. The matrix $\hat A$ is numerical and contains the transition rates: the coefficient $\hat A_{ij}$ is equal to the rate of the transition $j \to i$. As the only allowed transitions are from $j$ to $j, j\pm 1$, the matrix $\hat A$ is tridiagonal; in addition all its column sums are equal to 0, $\sum_{i=0}^n \, \hat A_{ij} = 0$ (a consequence of the conservation law $\sum_{i=0}^n P_i(t) = 0$). The last column of $\hat A$ is also zero since $n$ is an absorbing state. For this reason, it is sufficient to consider the restriction $A$ of $\hat A$ to its first $n$ rows and $n$ columns, $A = (\hat A_{ij})_{0 \leq i,j \leq n-1}$. The differential system can then be written as 
\be
{{\rm d} \over {\rm d}t} \, \vec P(t) = A \vec P(t)\,, \qquad \hbox{where\ }\vec P(t) = (P_i(t))_{0 \leq i \leq n-1},
\ee 
supplemented by the extra equation ${{\rm d} \over {\rm d}t} P_n(t) = k^+_{n-1} P_{n-1}(t)$, as well as the initial condition $\vec P(0) = (1,0,\ldots,0)^t$.

Our purpose is to compute 
\be
\la T \ra = \int_0^\infty {\rm d}t \: t\,{{\rm d}P_n(t) \over {\rm d}t}  = k^+_{n-1} \int_0^\infty {\rm d}t \: t\,P_{n-1}(t).
\ee
Up to the factor $k_{n-1}^+$, this is the last component of the length $n$ vector $\la \vec T \ra = \int_0^\infty {\rm d}t \: t\,\vec P(t)$.

Let us observe that the solution of the differential system is formally given by 
\be
\vec P(t) = \sum_{k=0}^\infty {t^k \over k!} \, A^k \vec P(0) = {\rm e}^{tA} \, \vec P(0),
\ee
where, as indicated, the exponential of the matrix $tA$ is defined by its Taylor series. Applying $A^2$ on $\la \vec T \ra$, using $A^2 \vec P(t) = {{\rm d}^2 \over {\rm d}t^2} \vec P(t)$ and integrating by parts, we obtain
\be
A^2 \la \vec T \ra = \int_0^\infty {\rm d}t \: t\,{{\rm d}^2 \vec P(t) \over {\rm d}t^2} = \Big[t\,{{\rm d} \vec P(t) \over {\rm d}t}\Big]_0^\infty - \Big[\vec P(t)\Big]_0^\infty = \vec P(0),
\ee
where we have used the fact that $\vec P(t)$ goes exponentially to zero when $t \to \infty$. 
From this we obtain that $\la \vec T \ra = A^{-2} \vec P(0)$. Taking the last component of this vector equation and using $\vec P(0) = (1,0,0,\ldots,0)$, we find that
\be
\langle T \rangle = k_{n-1}^+ \, (A^{-2})_{n-1,0} = k_{n-1}^+ \, \sum_{i=0}^{n-1} \, A^{-1}_{n-1,i} \, A^{-1}_{i,0}
\ee
is proportional to the scalar product of the last row of $A^{-1}$ and its first column.

Let us first apply this formula to the simplest situation, namely when all transition rates $k^\pm_i$ are equal to 1. In this case, the entries of $A$ are equal to $+1$ on the lower and upper diagonals, namely $A_{i,i\pm 1}=+1$. They are equal to $-2$ on the main diagonal, with the exception of $A_{00}=-1$. One can verify that the inverse of $A$ is given by
\be
A^{-1}_{ij} = \max(i,j) - n, \qquad \quad 0 \leq i,j \leq n-1.
\ee
We therefore have $A^{-1}_{n-1,i} = -1$ and $A^{-1}_{i,0} = i-n$, and the following result for $\la T \ra$,
\be
\la T \ra = \sum_{i=0}^{n-1} \, (n-i) = {n(n+1) \over 2}.
\ee
\end{multicols}
In the completely general case with arbitrary rates, the matrix $A$ takes the following form
\be
A = \begin{pmatrix}
-k^+_0 & k^-_1 & 0 & \ldots & \ldots & 0 \cr
k^+_0 & -(k^+_1+k^-_1) & k^-_2 & \ldots & \ldots & 0 \cr
0 & k^+_1 & -(k^+_2+k^-_2) & \ldots & \ldots & 0 \cr 
0 & 0 & k^+_2 & \ldots & \ldots & 0 \cr
\ldots & \ldots & \ldots & \ldots & \ldots & k^-_{n-1} \cr
0 & 0 & 0 & 0 & k^+_{n-2} & -(k^+_{n-1}+k^-_{n-1})
\end{pmatrix}.
\ee
\begin{multicols}{2}

Because the column sums of $A$ are all zero except the last one which is $-k^+_{n-1}$, the last row of $A^{-1}$ is constant and equal to $A^{-1}_{n-1,i} = -{1 \over k^+_{n-1}}$. Beside, the first column of $A^{-1}$ is somewhat more complicated and given by 
\be
A^{-1}_{i,0} = -{1 \over k^+_i} \; \sum_{\ell=0}^{n-1-i} \: {k^-_{i+1} \ldots k^-_{i+\ell} \over k^+_{i+1} \ldots k^+_{i+\ell}}, \qquad i=0,1,\ldots,n-1,
\ee
where, by convention, the term $\ell=0$ is set to 1. We then obtain
\be
\langle T \rangle = \sum_{i=0}^{n-1} \: {1 \over k^+_i} \; \sum_{\ell=0}^{n-1-i} \: {k^-_{i+1} \ldots k^-_{i+\ell} \over k^+_{i+1} \ldots k^+_{i+\ell}},
\ee
which is equivalent to (\ref{T}) upon the interchange of the two summations.

\subsection{Testing another example: the 3-step enzymatic reaction}

To validate the above recipes, let us apply them to a slightly more complicated (3-step) enzymatic reaction shown in Fig.7. 

\begin{center}
\includegraphics[width=8cm]{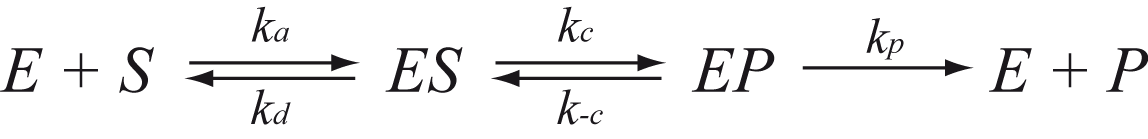} \\
\end{center}
\begin{small} \textbf{Figure 7.} Extended MM reaction where the catalytic reaction is reversible. \end{small}\\

The 3-step reaction is more realistic than the traditional 2-step MM reaction of Fig.2. Indeed, the last transition of the MM scheme (rate $ k_{c} $ in Fig.2) mixes two elementary reactions: the catalysis \textit{sensu-stricto} and the dissociation of the product from the enzyme. From a chemical viewpoint, the catalytic reaction has no obvious reason to be micro-irreversible. Instead, irreversibility can be understood as the consequence of the very low concentration of the product in the medium preventing it to rebind to the enzyme. This is true \textit{in vitro} when measuring initial reaction rates because no product molecules are added in the reaction mixture. This is also generally true \textit{in vivo} because the products are immediately removed by sequestration or by subsequent reactions of which they are the substrates, so that their steady state concentration remains negligible in the cell. Whatever the method used, the rate corresponding to the scheme of Fig.7 is the following function of the substrate concentration

\begin{equation} 
k = \frac{k_{c}k_{p}k_{a}[S]}{k_{p}k_{c}+k_{p}k_{d}+k_{d}k_{-c}+(k_{p}+k_{c}+k_{-c})k_{a}[S]}.
\end{equation}
The most direct methods to obtain this result are listed below.

\noindent
\subsubsection{The King and Altman method}

\medskip
\leftskip=9mm
\noindent
The first order enzyme cycle corresponding to Fig.7 is represented in Fig.8.

\begin{center}
\includegraphics[width=2.5cm]{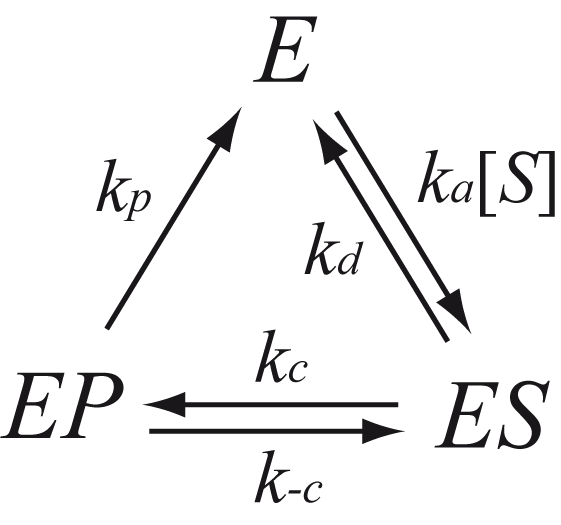} \\
\end{center}
\begin{small} \textbf{Figure 8.} The first order cycle of the reaction of Fig.7, whose drawing is a prerequisite for using the King and Altman procedure. \end{small}\\

The turnover rate of a single-enzyme corresponds to the reaction rate and is
\begin{equation}
k = k_{p}P_{EP},
\end{equation}
where $ P_{EP} $ is given by the graphical method of King and Altman.

\begin{center}
\includegraphics[width=7cm]{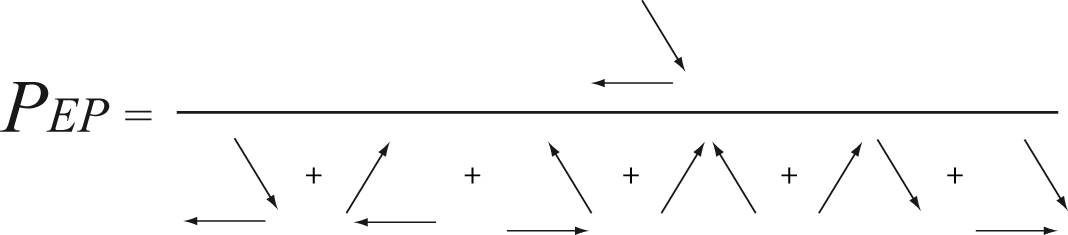} \\
\end{center}

$P_{EP}$ can be calculated by hand by replacing every graph by the product of rates corresponding to the arrows, but the result can be more safely obtained using an algorithm such as the KAPattern algorithm of \cite{Beard}, very useful for more complex networks.

\leftskip=0cm

\noindent
\subsubsection{The sum of direct conversion times}

\medskip
\leftskip=9mm
\noindent
Here the rate reads
\begin{equation} 
\frac{1}{k} = \frac{1}{k_{p}} + \frac{1}{k_{c}\frac{k_{p}}{k_{-c}+k_{p}}} + \frac{1}{k_{a}[S]\frac{k_{c}\frac{k_{p}}{k_{-c}+k_{p}}}{k_{d}+k_{c}\frac{k_{p}}{k_{-c}+k_{p}}}}.
\end{equation}

\leftskip=0cm

\noindent
\subsubsection{Markovian modeling of a random walk}

\medskip
\leftskip=9mm
\noindent
We obtain by this method
\begin{equation} 
\frac{1}{k} = \frac{1}{k_{a}[S]} + \frac{1}{k_{c}}+ \frac{1}{k_{p}} + \frac{k_{d}}{k_{a}[S] k_{c}} + \frac{k_{-c}}{k_{c}k_{p}} + \frac{k_{d}k_{-c}}{k_{a}[S]k_{c}k_{p}}.
\end{equation}

\leftskip=0cm

\noindent
Acknowledgement P.R. is Senior Research Associate of the Belgian Fonds National de la Recherche Scientifique (FRS-FNRS).

\end{multicols}


\begin{thebibliography}{19}
\bibitem{michel} D.  Michel, Biochimie 91, 933 (2009)
\bibitem{Cha} S. Cha, J. Biol. Chem. 243, 820 (1968)
\bibitem{Ninio1986} J. Ninio, In Accuracy in Molecular Processes, Kirkwood. T.B.L., R. Rosenberger, and D. J. Galas, eds., pp. 291-328 (London, Chapman \& Hall, 1986)
\bibitem{Ninio1987} J. Ninio, Proc. Nat. Acad. Sci. USA 84, 663 (1987)
\bibitem{Michaelis} L. Michaelis, M. Menten, Biochemistry Zeitung 49, 333 (1913)
\bibitem{Biot} J.-B. Biot, Mem. Acad. Sci. 15, 93 (1838)
\bibitem{Dhatt} S. Dhatt, K. Bhattacharyya, J. Math. Chem. 51, 1467 (2013).
\bibitem{Briggs} G.E. Briggs, J. B. Haldane, Biochem. J. 19, 338 (1925)
\bibitem{King-Altman} E.L. King, C.A. Altman, J. Phys. Chem. 60, 1375 (1956)
\bibitem{Qian} H. Qian, Biophys. J. 95, 10 (2008)
\bibitem{Fisher} J.R. Fisher, V.D. Hoagland Jr, Adv. Biol. Med. Phys. 12, 163 (1968)
\bibitem{Knorre} D.G. Knorre, E.G. Malygin, Dokl. Akad. Nauk SSSR 207, 1391 (1972)
\bibitem{Malygin} E.G. Malygin, G.A. Hattman, J. Theor. Biol. 242, 627 (2006)
\bibitem{Xie1998} H.P. Lu, L. Xun, X.S. Xie, Science 282, 1877 (1998)
\bibitem{Doussal} P. Le Doussal, Phys. Rev. Let. 62, 3097 (1989)
\bibitem{Zhou} Y. Zhou, J.E. Pearson, A. Auerbach, Biophys. J. 89, 3680 (2005)
\bibitem{Codling} E.A. Codling, M.J. Plank, S. Benhamou, J. R. Soc. Interface 5, 813 (2008)
\bibitem{Qian2} H. Qian, L.M. Bishop, Int. J. Mol. Sci. 11, 3472 (2010)
\bibitem{Beard} F. Qi, R.K. Dash, Y. Han, D.A. Beard, BMC Bioinformatics doi: 10.1186/1471-2105-10-238 (2009)

\end{thebibliography}
\end{document}